# Intrinsic Bistability

# In Quantum Point Contacts with in-plane Side Gates

J. Charles[1], M. Cahay[1,2], and R. S. Newrock[2]

[1] School of Electronics and Computing Systems
University of Cincinnati, Cincinnati, Ohio 45221, USA

[2] Physics Department, University of Cincinnati, Cincinnati, Ohio 45221, USA

## Abstract

We study the onset of intrinsic bistability and accompanying hysteresis in a single quantum point contact (QPC) with in-plane side gates in the presence of lateral spin-orbit coupling. The hysteresis in the conductance versus common gate voltage applied to the two side gates exists only if the narrow portion of the QPC is long enough. The hysteresis is absent if the effects of electron-electron interaction are neglected but increases with the strength of the electron-electron interaction. The hysteresis appears in the region of conductance anomalies, i.e., less than $2e^2/h$, and is due to multistable spin textures in these regions.

Over the last ten years, there have been many experimental reports of anomalies in the quantized conductance of QPCs and there is a growing consensus that these conductance anomalies are indirect evidence for the onset of spin polarization in the QPC's channel [1]. These anomalies include conductance plateaus around $0.5 \times 2e^2/h$ and $0.7 \times 2e^2/h$ [1]. There have been some recent reports that the number and location of these conductance anomalies can be tuned by deliberately introducing a broken symmetry in the QPC's electrostatic confining potential [2-5].

Shailos et al. investigated the linear transport properties of QPCs whose symmetry was deliberately broken in a controlled manner with two sets of gates [2]. In their QPCs, a conventional split-gate is used in conjunction with an additional perturbing gate or finger gate whose action is to modulate the electron density on *one* side of the device. As the voltage applied to this additional finger gate was varied, Shailos et al. observed several reproducible conductance anomalies below the last integer plateau, in addition to strong modifications to the integer-plateau staircase. The conductance curves were characterized by a marked hysteresis as the voltage on the split-gate was swept in opposite directions. There are many extrinsic mechanisms that could be responsible for the observed hysteresis, including the charging or discharging of surface states at the metal–semiconductor junction or of impurity states within the heterostructure close to the two-dimensional electron gas (2DEG) in the vicinity of the narrow portion of the QPC. Shailos et al. pointed out that the sensitivity of the hysteresis to device geometries and biasing conditions could be used to assess the microscopic configuration changes in the QPC and their connection with the observed conductance anomalies [2].

Ihnatenska and Zozoulenko calculated the conductance of a conventional QPC as a function of a symmetric potential applied to a split-gate in the framework of the density functional theory in the local density approximation (LDA) [6]. They showed that the spin degeneracy of the conductance channels is lifted leading to a broad plateau near ~ 0.5 $G_0$. Their calculated conductance curves show a hysteresis for forward and reverse sweeps of the applied gate voltage to the split-gate. Ihnatenska and Zozoulenko attributed this feature to the formation of weakly coupled quasi boundstates (magnetic impurities) inside the QPC. However, Ihnatsenka and Zozoulenko stressed that the formation of magnetic impurities in the narrow portion of the QPC might be an artifact of the LDA when localization of charge occurs in the device.

The goal of the present paper is to investigate the conditions under which multiple hysteresis curves and accompanying intrinsic bistability (IB) can be observed in the conductance of QPCs with in-plane side gates in the presence of *lateral spin-orbit coupling* (LSOC). We show that IB regions are most prominent near the conductance anomalies, i.e., less than $2e^2/h$, and are linked to a non-zero spin polarization and the availability of mutistable spin polarized states in these regions.

Recently, we showed that LSOC in InAs/InAlAs and GaAs/AlGaAs QPCs with in-plane side gates can be used to create a strongly spin-polarized current by purely electrical means *in the absence of any applied magnetic field* [7-10]. The use of a Non Equilibrium Green's Function (NEGF) approach to calculate the conductance of QPCs shows that the onset of spin polarization in devices with in-plane side gates required three ingredients: (1) an asymmetric lateral confinement, (2) a LSOC induced by lateral confining potential of the QPC, and (3) a strong electron-electron interaction [11,12]. The NEGF approach was used to study the ballistic conductance of asymmetrically biased side-gated QPCs in the presence of LSOC and electron-electron interaction for a wide range of QPC dimensions and gate bias voltage [12].

The model QPC we have used is shown in Fig. 1, where the white region represents the QPC channel with openings at the ends. In Fig.1, the gray area represents the etched isolation trenches that define the lithographic dimensions of the QPC constriction. The black strips show the four contact electrodes connected to the QPC device: source, drain and two side gates. Symmetric and asymmetric side gate voltages can be applied. We consider the QPC in Fig. 1 to be made from a nominally symmetric GaAs quantum well (QW). Spatial inversion asymmetry is therefore assumed to be negligible along the growth axis (*z* axis) of the QW and the corresponding Rashba spin-orbit interaction is neglected. The Dresselhaus spin-orbit interaction due to the bulk inversion asymmetry in the direction of current flow was also neglected. The only spin-orbit interaction considered is the LSOC due to the lateral confinement of the QPC channel, provided by the isolation trenches and the bias voltages of the four side gates [7,10]. The free-electron Hamiltonian of the QPC is given by

$$H = H_0 + H_{SO}$$
$$H_0 = \frac{\hbar^2(k_x^2 + k_y^2)}{2m^*} + U(x, y) \qquad (1)$$
$$H_{SO} = \beta\vec{\sigma}\cdot(\vec{k}_x \times \vec{\nabla}U(x,y))$$

In equation (1), $H_{SO}$ is the LSOC interaction term, $\beta$ the intrinsic spin-orbit coupling parameter, $\vec{\sigma}$ the vector of Pauli spin matrices, and $\vec{B}_{SO} = \beta(\vec{k}_x \times \vec{\nabla}U(x,y))$ is the effective magnetic field induced by LSOC. The effective mass in the InAs channel was set equal to $m^* = 0.023m_0$, where $m_0$ is the free electron mass. The 2DEG is assumed to be located in the (*x, y*) plane, *x* being the direction of current flow from source to drain and *y* the direction of transverse confinement of the channel. $U(x, y)$ is the confinement potential, which includes the potential introduced by the four side gates and the conduction band discontinuity at the GaAs/vacuum interface.

The conductance through the QPC was calculated using a NEGF method under the assumption of ballistic transport [11,12]. We used a Hartree-Fock approximation following Lassl et al. [13] to include the effects of electron-electron interaction. The latter was taken into account by considering a repulsive Coulomb contact potential, $V_{int}(x,y;x',y') = \gamma \delta(x-x') \delta(y-y')$, where $\gamma$ indicates the electron-electron interaction strength. This lead to the addition of an interaction self-energy, $\Sigma_{int}^{\sigma}(x,y)$, to the Hamiltonian in Eq.(1). At the interface between the narrow portion of the QPC and vacuum, the conduction band discontinuities at the bottom and the top interface were modeled, respectively, as

$$\Delta E_c(y) = \frac{\Delta E_c}{2}\left[1+\cos\frac{\pi}{d}\left(y-\frac{w_1-w_2}{2}\right)\right] \qquad (2)$$

and

$$\Delta E_c(y) = \frac{\Delta E_c}{2}\left[1+\cos\frac{\pi}{d}\left(\frac{w_1+w_2}{2}-y\right)\right] \qquad (3)$$

to achieve a smooth conductance band change; $d$ was selected to be in the nm range to represent a gradual variation of the conduction band profile from inside of the QPC to the vacuum region. A similar grading was also used along the walls going from the wider part of the channel to the central constriction of the QPC (Fig.1). This gradual change in $\Delta E_c(y)$ is responsible for the LSOC that triggers the spin polarization of the QPC in the presence of an asymmetric potential between the side gates. In our simulations, the parameter $d$ appearing in Eqns. (2) and (3) was set equal to 1.6 nm. The conductance of the QPC was then calculated using the NEGF with a non-uniform grid configuration containing more grid points at the interface of the QPC with vacuum. All calculations were performed at a temperature $T = 4.2\ K$.

Figure 2 is a plot of the conductance G (in units of $e^2/h$) of a InAs QPC as a function of the common mode signal $V_{sweep}$ applied to the two in-plane side gates. The biasing conditions on the gates are $V_{sg1}$ = -0.2V + $V_{sweep}$ and $V_{sg2}$ = 0.2V + $V_{sweep}$. The temperature is set equal to 4.2K and the device dimensions are $l_1 = l_2$ + 32nm, $w_2$ = 16nm, and $w_1$ = 48nm. The following parameters were used. $V_{ds}$ = 0.3mV, T = 4.2K, $\gamma$ = 3.7 in units of $\hbar^2/2m^*$, and $\beta$ = 200 Å$^2$. The conductance is calculated for two different lengths $l_2$ of the narrow portion of the QPC, $l_2$ = 24 and 48 nm. The solid and dashed curves correspond to the forward and reverse sweeps, respectively. Figure 2 indicates that there is no hysteresis in the conductance curves of the QPC with shorter length ($l_2$ = 24 nm) and there is no conductance anomaly below $2e^2/h$ in that case. For $l_2$ = 48 nm, there are two hysteresis loops in the conductance curves around anomalous conductance plateaus located near 0.5x$2e^2/h$ for the forward sweep and 0.5x$2e^2/h$ and 0.75x$2e^2/h$ for the reverse sweep. We found that the hysteresis loop starts to be prominent when the QPC channel length $l_2$ is above 30nm and appears in conjunction with the development of a conductance anomaly below the normal conductance plateau. The size of the hysteresis loop increases beyond that threshold as the 0.5x$2e^2/h$ anomaly becomes more pronounced [12]. The conductance anomalies are linked to the presence of various spin textures and non-zero spin polarization in the narrow portion of the QPC [12,14-17]. Since in the region of the conductance anomalies, there is substantial difference between the conductance curves for the forward and reverse sweeps, the hysteresis is an indirect proof of the existence of a finite spin polarization in the narrow portion of the QPC. In fact, even though not shown here, the conductance spin polarization $\alpha = [G_\uparrow - G_\downarrow]/[G_\uparrow + G_\downarrow]$ (where $G_\uparrow$ and $G_\downarrow$ are the conductance due to the majority and minority spin bands, respectively) was found to be non-zero in the negative differential resistance (NDR) regions and was close to unity for both the forward and reverse sweeps. Figure 2 also shows the results of the conductance calculations for $l_2 = 48$ nm with $\gamma = 0.0$, i.e., when the effects of electron-electron interaction are neglected. There is no hysteresis and no anomalies in the conductance plots in that case. This is in agreement with our earlier findings that a net spin polarization in asymmetrically biased QPC in the presence of LSOC can only be observed when there is sufficient electron-electron interaction in the device [7,11]. The hysteresis shown in Fig.2 is similar to the one observed in the NDR regions of resonant tunneling devices [18-22] and is linked to the feedback of space-charge effects in the narrow portion of the QPC in the presence of LSOC. The NDR regions in the forward and reverse sweeps are linked to multiple reflections between the edges of the narrow portion of the QPC [11,12].

Figure 3 is a plot of the spin density $n_\uparrow (x,y) - n_\downarrow (x,y)$ (in units of m$^{-2}$) calculated in the vicinity of the maxima (points A and B for the reverse and forward sweeps, respectively) and minima (points C and D for the reverse and forward sweeps, respectively) of the rightmost hysteresis loop in Fig.2. As shown in Fig.3, there is a substantial difference between the spin textures for the forward and reverse sweeps (compare the spin textures at points A and B, and C and D). This figure shows that hysteresis in the conductance curves results from the presence of metastable spin configurations for biasing conditions corresponding to conductance anomalies. The maximum in spin density at points A and C on the reverse sweep is about 3 orders of magnitude smaller than at points B and D in the forward sweep. Points A and B correspond to the maximum in the conductance spin polarization $\alpha$ for the reverse and forward sweep, respectively.

There is an emerging consensus in the spin origin of the conductance anomalies observed in QPCs of various geometries [1]. Many theories of conductance anomalies have been proposed based on that premise, including a spontaneous spin polarization in the narrow portion of the QPC resulting from the exchange-correlation interaction [16,17,23], the formation of quasibound states in the narrow portion of the QPC [24], and some based on the Kondo effect [25]. Our NEGF approach show that hysteresis in the conductance curves of a single QPC with in-plane side gates in the presence of LSOC is absent if the narrow portion of the QPC is not long enough or if the effects of electron-electron interaction are neglected. The hysteresis appears in the region of conductance anomalies, i.e., less than $2e^2/h$, and is linked to the existence of multistable spin textures in these regions. The presence of IB in the conductance of asymmetrically biased QPC structures and their appearance near conductance anomalies constitute another indirect proof of the onset of spin polarization in the devices. However, the IB discussed here could be masked by other sources of hysteresis such as the presence of dangling bonds on the side walls of the QPC [10].

Since asymmetry in the QPC design or biasing conditions is required for the observation of IB regions, the latter should also be observable in QPCs with independently biased top gates. An understanding of the number, size, temperature and bias sensitivity of the IB regions in the conductance curves of asymmetrically biased QPCs should provide a new challenge to the many theories developed to date to understand the conductance anomalies in QPCs. A thorough study of the size of IB regions could be used to refine and champion one of the existing theories of the microscopic origin of the conductance anomalies [1].

**Acknowledgment**

This work is supported by NSF Award ECCS 1028423. James Charles acknowledges support under NSF-REU award 007081.

## Figure Captions

**Fig. 1:** Schematic illustration of the QPC geometrical layout used in the numerical simulations.

**Fig. 2:** Conductance G (in units of $G_0$) of an InAs QPC as a function of the common mode signal $V_{sweep}$ applied to the two in-plane side gates. The following biasing conditions are used: $V_{ds}$ = 0.3mV, $V_{sg1}$ =- 0.2V + $V_{sweep}$ and $V_{sg2}$ = 0.2V + $V_{sweep}$. The temperature is set equal to 4.2K and the device dimensions are $l_2$ = 48nm, $l_1$ = 32nm, $l_1 = l_2$ + 32nm, $w_2$ = 16nm, and $w_1$ = 48nm. $\gamma$ = 3.7 in units of $\hbar^2/2m^*$ and $\beta$ = 200 Å$^2$. The solid and dashed curves correspond to the forward and reverse sweeps, respectively. Also shown for comparison are the conductance plots for a QPC with $l_2$ = 24nm, which show no hysteresis in that case. Also shown is the conductance plot for $l_2$ = 48 nm with $\gamma$ = 0.0, i.e., when the effects of electron-electron interaction are neglected.

**Fig. 3:** Plot of spin density $n_\uparrow (x,y) - n_\downarrow (x,y)$ (in units of m$^{-2}$) in the QPC channel at $V_{sweep}$ = 0 mV for both forward and reverse sweeps. In this simulation, $V_{sg1}$ = -0.2V + $V_{sweep}$ and $V_{sg2}$ = 0.2V + $V_{sweep}$. The temperature is set equal to 4.2K and the device dimensions are $l_2$ = 48nm, $l_1 = l_2$ + 32nm, $w_2$ = 16nm, and $w_1$ = 48nm. The other parameters are: $V_{ds}$ = 0.3mV, T = 4.2K, $\gamma$ = 3.7 in units of $\hbar^2/2m^*$, and $\beta$ = 200 Å$^2$. The different plots correspond to the spin density in the QPC calculated (going clockwise from the upper left frame) at points A, C, D, and B in Fig.2, respectively.

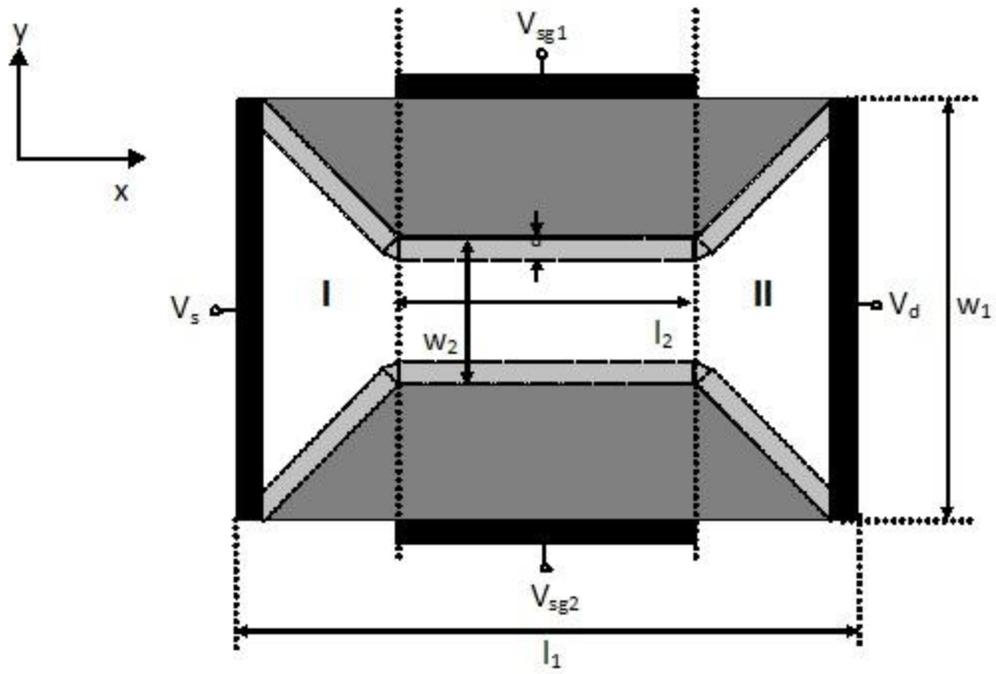

**Figure 1**

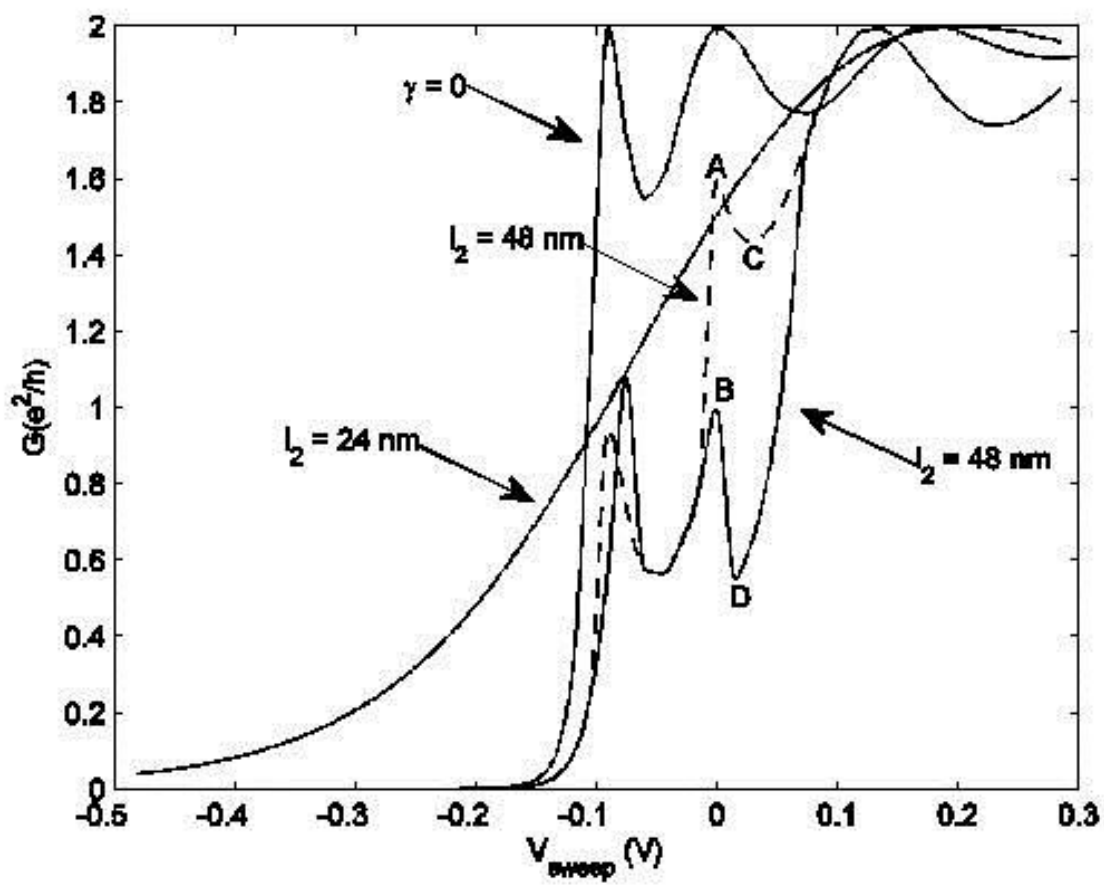

**Figure 2**

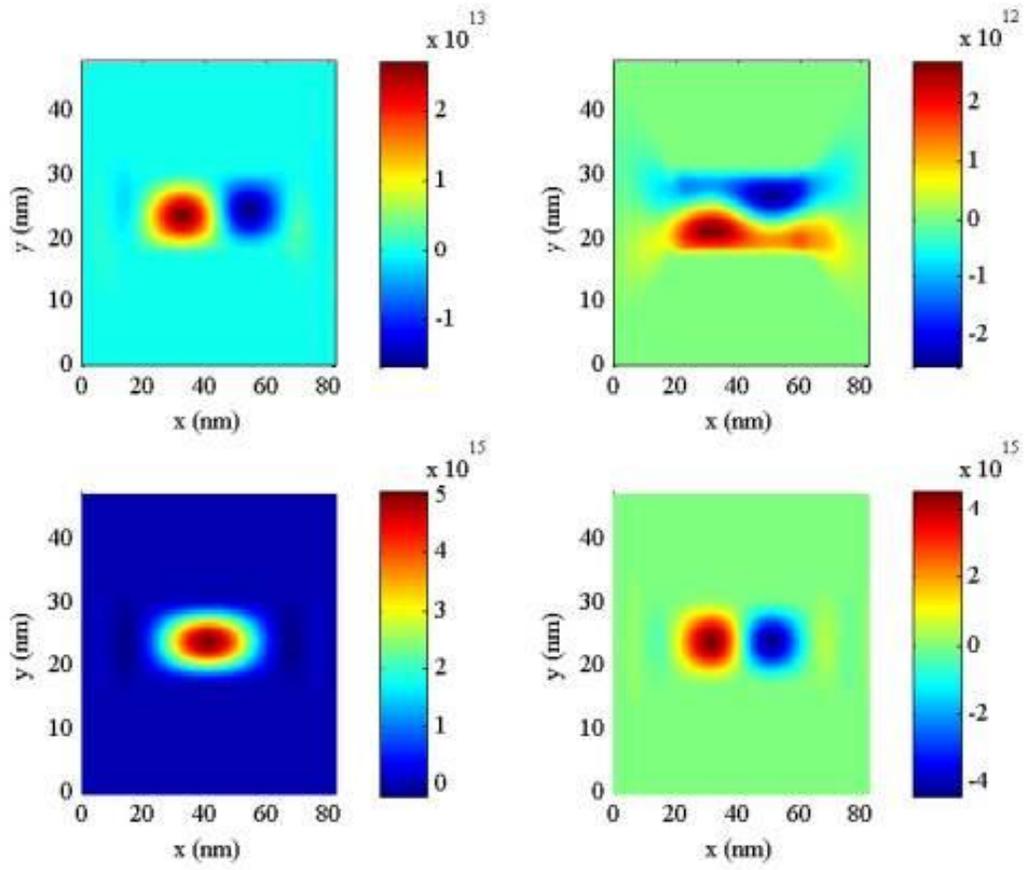

**Figure 3**